# Noise and Transport Characterization of Single Molecular Break Junctions with Individual Molecule.

V.A. Sydoruk,[1,⊥] D. Xiang,[1,⊥] S.A. Vitusevich,[1*] M.V. Petrychuk,[2] A.Vladyka[1], Y. Zhang,[1] A. Offenhäusser,[1] V.A. Kochelap,[3] A.E. Belyaev,[3] and D. Mayer[1]

[1]*Peter Grünberg Institut, Forschungszentrum Jülich, D-52425, Germany.*
[2]*Taras Shevchenko National University of Kyiv, 03022, Ukraine.*
[3]*Institute of Semiconductor Physics, NASU, Kyiv, 03028, Ukraine.*
[⊥]*These authors contributed equally to this work.*

**Abstract.** We studied the noise spectra of molecule-free and molecule-containing mechanically controllable break junctions. Both types of junctions revealed typical $1/f$ noise characteristics at different distances between the contacts with square dependence of current noise power spectral density on current. Additional Lorentzian-shape ($1/f^2$) noise components were recorded only when nanoelectrodes were bridged by individual 1,4-benzenediamine molecule. The characteristic frequency of the revealed $1/f^2$ noise related to a single bridging molecule correlates with the lock-in current amplitudes. The recorded behavior of Lorentzian-shape noise component as a function of current is interpreted as the manifestation of a dynamic reconfiguration of molecular coupling to the metal electrodes. We propose a phenomenological model that correlates the charge transport via a single molecule with the reconfiguration of its coupling to the metal electrodes. Experimentally obtained results are in good agreement with theoretical ones and indicate that coupling between the molecule metal electrodes is important aspect that should be taken into account.

## I. INTRODUCTION

Theoretical and experimental investigations of charge transport through organic molecules attract increasing attention driven by the interest in fundamental aspects of charge transfer mechanisms and the vision of future applications in molecular electronics. Starting from the first demonstration[1] of the transition from the tunneling regime to point contact in scanning tunneling microscopy setup, during the past decade, experimental and theoretical advances have yielded significant insights into electron transport in metal-molecule-metal junctions.[2–7] In particular, charge transfer investigations have become feasible for individual molecules by realizing different nanoelectrode configurations where the electrodes are separated by a gap of molecular dimensions.[8–14]

---

[*] Corresponding author. E-mail: s.vitusevich@fz-juelich.de. On leave from Institute of Semiconductor Physics, NASU, Kiev, Ukraine.

Break junction setups with tunable distance between the nanocontacts, such as scanning tunneling microscopy, atomic force microscopy, or mechanically controllable break junctions (MCBJs), make it possible to control the number of molecules electrically connected in the junction.[13,15,16] Particularly MCBJs make use of an attenuation factor to control the electrode distance with unique accuracy which allows to investigate noise properties of nanocontacts in different regimes from fully connected to tunneling. The noise characteristics of bare metal break junctions were previously studied in different transport regimes up to break of the MCBJ.[17] It was shown that measured $1/f$ noise had two different power dependences on resistance of junctions corresponding to the transition from diffusive to ballistic transport.

The random telegraph noise was observed above the $1/f$ noise in alkyl-based metal/multi-molecule/metal junctions at high bias voltages and was explained as trapping-detrapping process via localized energy states.[18] The deviation in power low of flicker noise from $1/f$ behavior may be used to probe for the local environment of a single molecule contact.[19] The noise study in a tunneling current across the bridge molecule may allow to reveal rectifying properties in the nonlinear regime.[20] $1/f$ noise can be used to study the protein dynamics that modulate channel conductance.[21] Noise measurements of the current through a single-molecule junction was used to gain further information about electron tunneling processes, including case of a poor overlap of local orbitals resulting in nonlinear amplification and a huge current bursts.[22] In addition, noise spectroscopy allows one to study special features of single molecule junction transport which are not accessible by standard current-voltage measurements. For example, shot noise measurements indicating different conduction channels in the conductance,[23,24] noise measurements revealing huge current bursts under a poor overlap of local orbitals,[25] etc.

In the present work, high-stability break junction devices were fabricated and employed to investigate current behavior and noise characteristics of metal/single-molecule/metal (MSM) junctions at room temperature and low bias voltages. In contrast to the molecule-free case, for nanoelectrodes bridged by an individual molecule, we registered a Lorentzian-shape noise component, in addition to $1/f$ noise. The same noise phenomenon was observed also for a set of other organic molecules. Here we report detailed study of the Lorentzian-type noise component for the 1,4-Benzenediamine molecule. The observed Lorentzian-type noise is interpreted as a manifestation of a dynamic reconfiguration of molecular coupling to the metal electrodes. We believe that observed phenomenon is of quite general character. The underlying physics can be understood as follows.

The Lorentzian-type noise component is observed for small frequencies. While all existing times characterizing the motion of the electrons and nuclei in a molecule and metal contacts under equilibrium, as well as relaxation times in contacts, are smaller by many orders of magnitude than characteristic times of the noise recorded. This allows us to suggest that the characteristic noise times are related to slow non-equilibrium processes induced by low currents in the system of nanoelectrode-molecule. Indeed, when a current flows through a molecule, its electron subsystem becomes polarized, which induces *small* structural/configuration changes. Thus, charge transfer and structural/configuration changes are coupled. Small currents determine smallness of the forces inducing these changes and large characteristic times of such a reconfiguration. It is known that a system with two sets of strongly distinct characteristic times can demonstrate the following specific relaxation:[26] during short relaxation times the system reaches so-called "incomplete equilibrium", then it slowly ("quasistationary") relaxes in the course of a longer period of time. According to general theory of time-dependent fluctuations in systems under "incomplete equilibrium", the spectral density of fluctuations of a slow degree of freedom is defined by a Lorentzian-type dependence.[26] Introducing a slow configuration coordinate responsible for molecule-nanoelectrode coupling, we develop a phenomenological model, that takes into account a correlation between charge transfer via a single molecule and structural changes in the coupling. Thus we interpret the observed Lorentzian-type noise as the manifestation of a system reconfiguration under the charge transfer via a single molecule. Obtained experimental results, particularly dependencies of the noise amplitude and the characteristic noise frequency on the current, support this interpretation.

The paper is organized as follows. The used break junction setup is described in Section II. Details of current and noise measurements, as well as separation of the flicker noise and the Lorentzian-shaped noise are presented in Sections III. Discussion of experiments, their interpretation and introduction of the appropriate noise model are given in Section IV. In Section V we briefly summarize obtained results.

## II. BREAK JUNCTION SETUP

MCBJ chips under study were specially designed to be highly-stable that is especially important for noise measurements. In some cases 50 minutes stability was achieved. It should be noted that reported in the literature stability does not exceed 3 minutes. MCBJ chips were fabricated on the basis of spring steel substrates with a length of 44 mm, a width of 12 mm and a thickness of 0.1mm. The typical view and fabrication process are shown in Fig.1. Five important steps were done to produce a very stable junction (Fig.1b). The first is to deposit an insulating layer. A polyimide (HD-4100, HD Microsystem) about 3 µm thick was spin coated on the substrate. After baking it at 200°C for 20 minutes, the substrate was annealed for one hour at 300 °C at a pressure of $10^{-3}$ mbar to improve the stability and structure of the layer. After that, the e-beam lithography process was done. It consists of three separate steps. First, 200 nm of the positive tone resist (PMMA, Polymethylmetacylate 649.04 from ALLRESIST) is spin coated onto the substrate and baked at 180°C for 2 minutes. Second, the electrode pattern is written by means of a Leica Vistec EBPG-5000 plus lithography System. Finally, a standard development procedure is applied by inserting the substrate into development solution (ALLESIST AR 600-55) for about 50 seconds, then the substrate was transferred into 2-propanol to stop the development. After the development, the resist layer serves as a mask for the metal deposition. At this step, 2 nm Ti and 40 nm Au are deposited on the substrate surface in a vacuum chamber by e-beam evaporation. Then, the sample is immersed in acetone for lift off. And in the final step, the polyimide is isotropically dry etched to obtain a suspended metal bridge. This is done by reactive ion etching (RIE) at the following conditions: 32 sccm of oxygen and 8 sccm of $CHF_3$ and a power of 100 W.

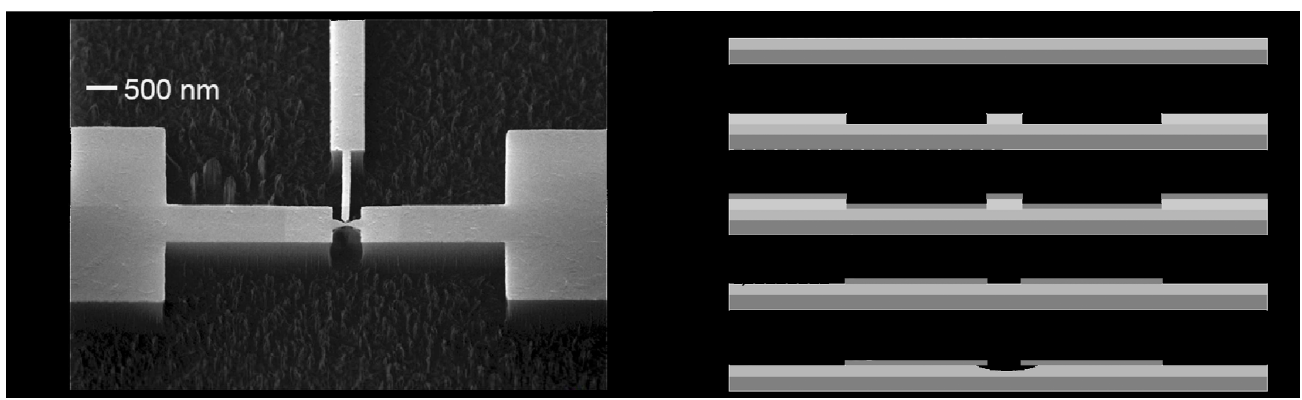

FIG. 1. (a) Scanning electron microscopy (SEM) image of a micro-fabricated MCBJ chip consisting of a freestanding metal bridge, with a central constriction, on top of an insulating spring steel substrate. A nano-gap appeared after the central constriction broke as the push rod applied a force to bend the substrate. (b) Chip fabrication process, which consist of five steps: spin coating isolate layer, e-beam lithography, gold deposition, lift off, etching.

The MCBJ chips were mounted into a home-made three-point bending apparatus. When the push rod exerts a bending force on the substrate, the movement in the Z direction causes an elongation of the constriction until the bridge breaks resulting in the formation of two separate nanoelectrodes which can be used to contact molecules.

Immediately before the break, the two electrodes are bridged by a few gold atoms. This stage demonstrates a typical discreteness in the junction conductance[9,11,15,25], $G = n\, G_0$, where $G_0 = 2e^2/h$ and $n$ is an integer (Fig.2). It is important that each quantized value of $G$ can be maintained for longer than two minutes in our setup, in some cases up to 50 minutes. The latter demonstrates that we have achieved very high mechanical stability.

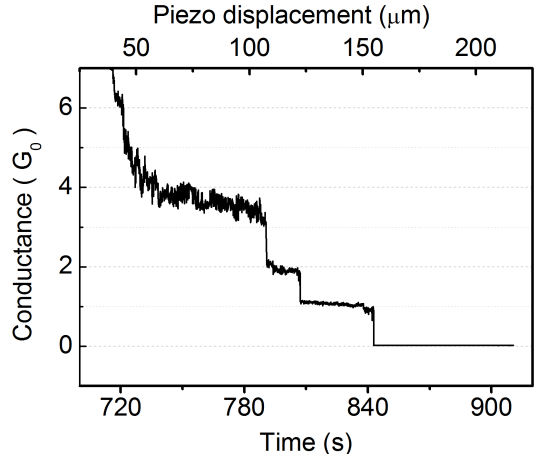

FIG. 2. The typical breaking process steps of pure gold MCBJ.

After breaking the bridge, electron transport occurs due to tunneling processes. The distance between the electrodes (the gap) for both opening and closing operation modes was tuned by bending or relaxing the substrate, respectively. Precision of the gap control is determined by the attenuation factor, $r$.[9] For the setup used, an attenuation factor of $r \approx 6\times10^{-6}$ was achieved. This implies that the gap between the electrodes can be controlled, in principle, with sub-angstrom accuracy.

### III. NOISE MEASUREMENT EXPERIMENTAL DETAILS

To reduce the 50Hz device noise usually registered in conventional units with a 220V source, a lead-acid battery was used to apply a defined bias to the samples. The samples were connected in series with load resistance, $R_{load}$. This scheme allows us to calculate the current through the sample and measure the voltage fluctuations. We measured the voltage on the sample, $V_S$, and the total voltage $V_M$ using a multi-meter. The current through the sample, $I_S$, was estimated by following formula:

$$I_S = (V_M - V_S)/R_{load} \qquad (1)$$

For all our measurement we used $R_{load} = 5\text{k}\Omega$. The noise signal from the sample was amplified by a home-made preamplifier with a gain of 177 and then by commercial amplifier (ITHACO 1201) with variable gain. The input impedance of the preamplifier in our measurement system can be assumed as 1MΩ at AC regime (above 0.3Hz). The intrinsic thermal noise of preamplifier and ITHACO amplifier are measured to be $2\times10^{-18}$ V$^2$Hz$^{-1}$ and $2\times10^{-17}$ V$^2$Hz$^{-1}$, respectively. The noise spectra were registered using a dynamic signal analyzer (HP35670A) and the data were transferred through GPIB interface to the computer.

The noise characteristics were measured in a broad frequency range, from 1Hz to $10^5$Hz. It should be noted, that the small value of parasitic capacitance caused by cables and junction itself was not higher than $C_P = 200$pF, therefore roll-off frequency of our noise set-up is 160kHz ($f_{roll\text{-}off} = 1/[2\pi C_P(R_{load}||R_{Sample})] \sim 1/(2\pi C_P R_{load}) \approx 160\text{kHz}$). Here $R_{sample}$ is the resistance of MCBJ under

investigation. The experimental studies were performed in vacuum about $10^{-4}$ mbar at room temperature (T~298K).

## IV. ELECTRIC CURRENT AND NOISE BEHAVIOR OF SINGLE MOLECULE

The noise characteristics of bare metal break junctions were previously investigated in detail from the diffusive to the ballistic transport regime.[17] Here, we focus our attention on investigating the noise properties of MCBJ samples in the tunneling regime. The molecule-free nanogap is used as a reference system for molecule-containing junctions. Voltage fluctuations were measured and Fourier transformed using a dynamic signal analyzer. The voltage power noise spectral density, $S_V$, measured over a bandwidth from 1 Hz to 100 kHz at different gap dimensions is shown in Fig.3. For small gap dimensions, the spectra follow exclusively $1/f^\alpha$-noise behavior ($0.9 < \alpha < 1.2$). For large gaps and high frequencies, the thermal noise of the nanojunctions dominates the $1/f$ noise. The total noise can be expressed by the following empirical equation:

$$S_V(f) = \frac{A}{f^\alpha} + 4k_B TR, \text{ with } \alpha \approx 1 \qquad (2)$$

Here, $A$ is the amplitude of $1/f$ noise at $f=1$Hz, $k_B$ is the Boltzmann constant, $T$ is the temperature, $R$ is the equivalent resistance, which consists of the nanojunction resistance and parallel-connected at AC regime load resistance, $R_{load} = 5$k$\Omega$ (specified by measurement setup). The first term of Eq. (2) describes the $1/f$ noise, while the second is frequency-independent thermal noise. With increasing gap size and decreasing tunneling current, $A$ decreases and $R$ increases up to the load resistance, which results in a dominant thermal noise component at high frequencies.

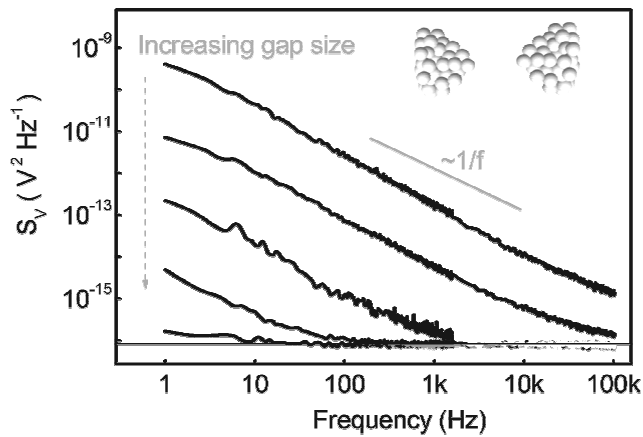

FIG. 3. The voltage noise power spectral density of the molecule-free junctions in the tunneling regime, measured with respect to different gap sizes between the gold nanoelectrodes. The tunneling resistance is measured to be 0.022, 0.030, 0.170, 1 and 10 MOhm from up to down, respectively. The fixed bias voltage applied to the junction is $V_B = 20$ mV. The horizontal black line represents the theoretically calculated thermal noise ($4kTR$) for $R$ equal to load resistance, $R_{Load} = 5$kOhm. The spectra show exclusively $1/f^\alpha$ ($0.9 \leq \alpha \leq 1.2$) noise behavior. The noise component decreases to the thermal noise level as the gap size is increased.

After investigation of noise characteristics of the molecule-free-gap junction, 1,4-benzenediamine (BDA), a molecule which contains two amine termini as binding groups, was integrated between the nanoelectrodes by a self-assembly process to generate a metal/single-molecule/metal junction. Therefore, a 1 mM ethanolic solution of 1,4-benzenediamine was prepared in a protective atmosphere where the oxygen level was less than 1 ppm. A 10 $\mu L$ droplet of this solution was placed on the junction area under nitrogen atmosphere.

The breaking process of the molecule modified nanowire was again followed by monitoring the conductance of the hetero-junction. From Fig.4a (black curve) one can observe the whole breaking process of the metal wire resulting in a sharp drop of the conductance after breaking the metal wire into two separated parts with terminal nanoelectrodes. After the sharp drop, a state with constant conductance (the lock-in state) was registered corresponding to bridging by the molecule under test. If continue the breaking process further, the second sharp drop will appear reflecting the complete break of the metal-molecules-metal junction. By decreasing the distance between the gold contacts, the lock-in state can be obtained again with the same value of conductance corresponding to the conductance of single molecule. Such a process (increasing and decreasing of electrode separation) can be repeated several times. The molecular junction finally transforms into the lock-in state which can be assigned to a configuration where the electrode gap is bridged by a single molecule (lowest conductance plateau). Due to the two amine groups of BDA, the molecule binds to both nanoelectrodes and a metal-molecules-metal junction is formed. The conductance of the junction in this device configuration is found to be equal $(6 \pm 1) \times 10^{-3}$ $G_0$ (Fig.4b). This value corresponds to the conductance of individual BDA molecule bridging the nanogap by establishing bonds to both electrodes.[25,27] Note, the single molecule conductance of BDA is well studied. Particularly, direct measurements of the single molecule conductance were compared with first-principles calculations[27] and excellent quantitative agreement was found. Additional argumentation of the single molecule bridging for the break junction setup used in this work can be found in Ref.[Cem.comm]

(Dropped by VA:The authors of Ref.[27] provide the first direct comparison between energy level alignment and single molecule transport measurements. The results of scanning tunneling microscope-based break-junction measurements of single molecule conductance were compared with first-principles calculations and the authors found excellent quantitative agreement.

The lock-in configurations can be obtained when the molecules were applied to the junction in opening operation mode, however with lower yield. In the opening operation mode the gap between the electrodes exceeds the size of the molecule. Under this condition, $1/f$ and thermal noise components were observed, similar to the case of molecule-free junctions. When the gap between the two separated electrodes became equal to the length of the molecule, the tunneling current suddenly jumped to the stable lock-in state, where the current became almost independent of the gap size. The typical plateaus at values below $1G_0$ are absent for molecule free junctions, Fig.4a. The conductance peak (Fig.4b) is located at $6\times10^{-3}$ $G_0$.

The *I-V* characteristics of MSM junction were continuously recorded in the lock in state (Fig.5). The dashed curve represents a fitting curve calculated from Simmons equation:[28]

$$J = \frac{e}{4\pi^2 \hbar d^2}\left\{(\phi_B - \frac{eV}{2})\cdot\exp\left[-\frac{2(2m)^{\frac{1}{2}}}{\hbar}\alpha(\phi_B - \frac{eV}{2})^{\frac{1}{2}}d\right] - (\phi_B + \frac{eV}{2})\cdot\exp\left[-\frac{2(2m)^{\frac{1}{2}}}{\hbar}\alpha(\phi_B + \frac{eV}{2})^{\frac{1}{2}}d\right]\right\}, \quad (3)$$

with the fitting parameters: $\alpha = 0.6$, $\phi_B = 1.0$ eV, $d = 0.82$ nm. Here $\phi_B$ is the energy gap between the Fermi level, $E_F$, of the gold electrodes and HOMO state of the molecule, $E_{HOMO}$. The obtained data correspond well with values earlier reported for BDA molecules. The analysis of $Ln(I/V^2)$ versus $1/V$ curves reveals that only direct tunneling mechanism was present within the bias range between -1 V and 1 V (insert to Fig.5). It should be noted, that recently a more rigorous model taking into account high voltage regime has been reported.[29] We studied relatively small voltage range, for which Simmons model can be used[30] to fit the experimental data.

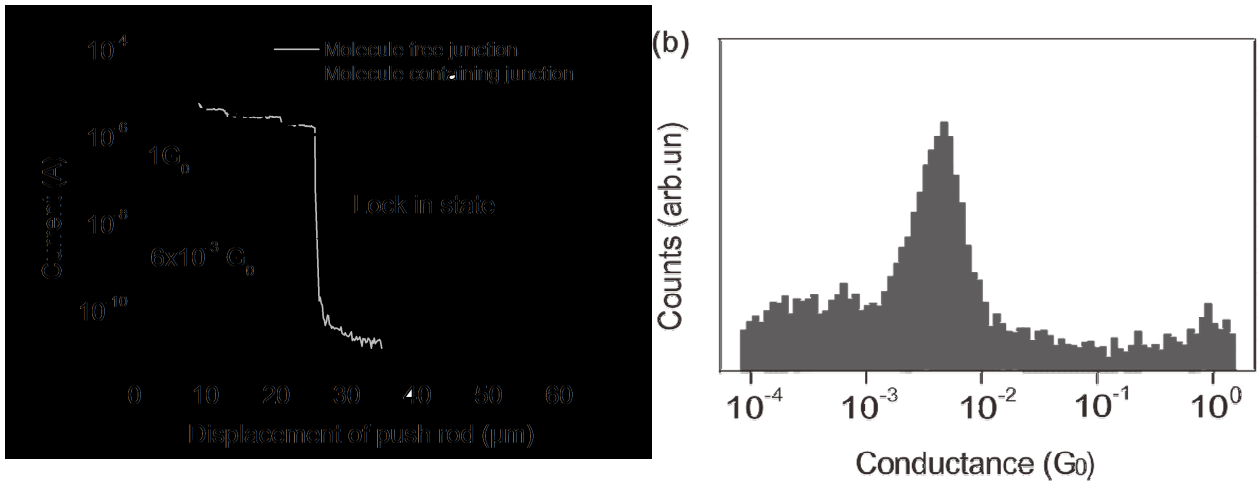

FIG. 4. (a) Representative tunnelling current responses of mechanically controlled break junctions as a function of push rod displacement in the gap opening period with an applied bias of 10mV. In the case of the molecule-containing junction (thick black curve), the current jumps to a lock-in state during the opening period. In the lock-in state, the current is almost independent of the gap size. In contrast, the lock-in state is absent in the molecule-free junction (thin grey curve). $G_0 = 2e^2/h$. The value $G = 6\times10^{-3}$ $G_0$ is a typical conductance for bridged 1,4-Benzenediamine molecule; (b) Conductance histogram of BDA molecular junctions. It was obtained using 100 curves of 5 different samples by repeating the opening and closing process.

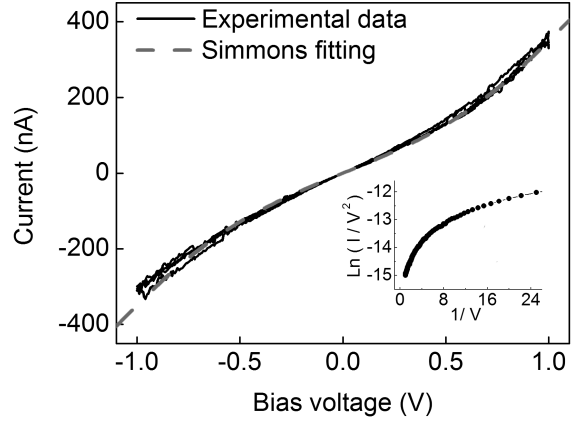

FIG.5 *I-V* characteristics of MSM junction continuously recorded in the lock in state (black curves) and Simmons fitting of the characteristics (dashed gray curve). Insert: Measured I-V characteristic plotted as Ln($I/V^2$) versus ($1/V$).

In this lock-in state, with a fixed gap, the noise characteristics were studied at the linear region of current-voltage characteristic with a bias of $V < 0.1$V (see Fig.6a). It should be emphasized that in addition to the $1/f$ noise characteristic for the molecule-free junction, a new noise component was revealed for the BDA molecule-bridged nanogap, as shown in Fig.6b. The new noise component was registered for all voltages studied. It has $1/f^2$ - frequency dependence, which is typical for a single Lorentz fluctuator. Now the noise characteristics can be described as the superposition of a $1/f$, a $1/f^2$ and thermal noise components:

$$S_V(f) = \frac{A}{f} + \frac{B/f_0}{1+(f/f_0)^2} + 4kTR \qquad (4)$$

Here, $B$ represents the amplitude of $1/f^2$ noise at low frequencies ($f \ll f_0$), $f_0$ is the so-called characteristic frequency derived from the inflection position of the total noise spectra. The value defined as $\tau_0 = 1/(2\pi f_0)$ characterizes time scale of a physical process responsible for $1/f^2$ fluctuations. Eq.(4) provides a perfect fitting model of the experimental data above 100 Hz, see Fig.6b. From fitted data, we determined a characteristic frequency of $f_0 = 600$ Hz and $\tau_0 = 0.25$ ms at a bias voltage of $V = 45$ mV. It should be noted that bump below 100Hz corresponds to second Lorentzian noise component. The component was disappeared in some voltage range due to shift to very low frequency range below the measurement frequency range.

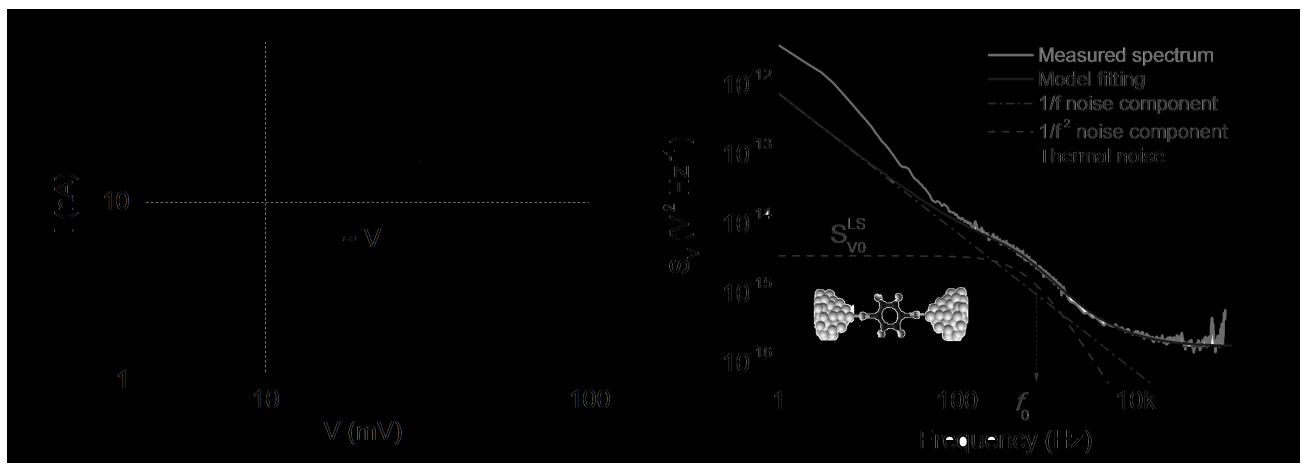

FIG. 6. (a) *I-V* characteristics of MSM junction recorded in the lock in state at low bias voltages $V < 0.1$ V; (b) The voltage noise power spectral density of the molecule-containing junction in the lock-in state, corresponding to a single BDA molecule bridging two electrodes measured at $V = 45$ mV. The solid gray curve is the measured noise density and the solid black curve presents the model fitting. The fitting curve is the superposition of $1/f$, $1/f^2$ and thermal noise components shown by a dash-dotted line, dashed curve, and thin gray line, respectively.

We measured noise characteristics attributed to charge transfer via individual molecule for stable lock-in state of the current over time. Each noise spectrum was obtained by averaging of 100 curves using a spectrum analyzer. For molecular junctions we measured about 50 MCBJs with the BDA molecule. And for each of the samples we measured more than 20 noise spectra at different applied voltages and gap sizes. Separating the $1/f$ noise of the gold electrodes and the thermal noise from the whole noise spectrum, we found a $1/f^2$ noise components that is specifically inherent in a single bridging molecule. This component was observed only in the case when molecule was coupled to both gold contacts. When one of the bonds was broken in the case of a bit increased distance between the contacts in the MSM junction the feature was not registered in the noise spectrum. Broken bond can be restored by a decreasing of the distance between contacts, and the $1/f^2$ component appears again.

For the flicker noise component, we found that the noise spectral density scales *quadratically* with the current in both molecule-free and molecule-containing junctions. For the $1/f^2$ noise component, *linear dependences* of the noise spectral density and the characteristic frequency on current were observed. The results are shown in Fig.7.

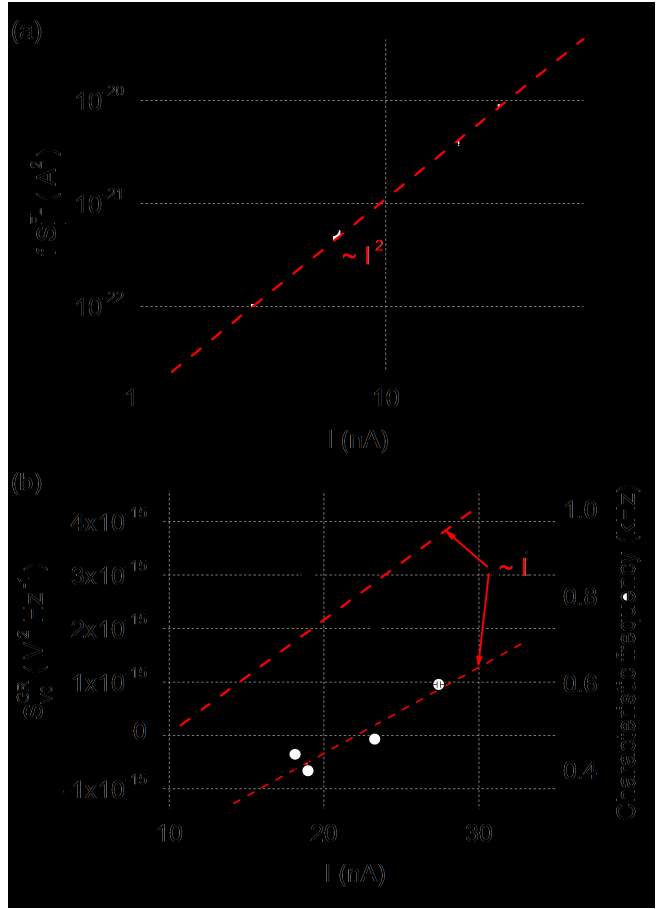

FIG. 7. (a) The measured at $f = 1$ Hz normalized power noise spectral density of the flicker noise component, $fS_I^{FL}$, has a quadratic dependence on current in molecule-containing as well as in molecule-free junctions; (b) Both the plateau of Lorentzian-shape noise component, $S_{V0}^{LS}$, and the characteristic frequency, $f_0$, linearly depend on current. Results are shown for a 1,4-Benzenediamine molecule bridging the nanogap of MCBJ at fixed gap size between two nanoelectrodes.

Below, we propose a phenomenological model that correlates the charge transport via a single molecule with reconfiguration of its coupling to the metal electrodes.

## IV. DISCUSSION

It is well known that $1/f^2$ noise may appear in mesoscopic structures, for example in field-effect transistors with submicron gate area.[31] This kind of the noise is generated due to random processes of electron trapping/detraping by a single defect. This mechanism is not likely to be applied to the case of a single molecule bridging the nanocontacts.[18] We suggest that the $1/f^2$ noise components recorded in the low-bias regime with low values of the characteristic frequencies $f_0$ can be understood as a result of a dynamic reconfiguration of molecular coupling to the metal electrodes.[32–35]

We start with discussion of the surprising fact of small frequencies of the observed $1/f^2$ noises. All existing times characterizing the motion of the electrons and nuclei in a molecule and metal contacts under equilibrium, as well as relaxation times in contacts, are smaller by many orders of magnitude than the characteristic noise times ($\sim 1/f_0$) recorded. We assume that large values of the characteristic noise times are related to non-equilibrium processes in the molecules induced by low currents. These processes can be considered as follows. When current flows through a molecule, the electron subsystem of the latter becomes polarized. The processes induce *small* structural/configuration changes. The charge transfer and structural/configuration changes are coupled. Small currents determine the smallness of the forces inducing the changes and, thus, large characteristic times of these changes. Let $X(t)$ be the configuration coordinate, generally dependent on time $t$. The junction resistance is taken to be a function of $X$: $R = R[X(t)]$. Then, fluctuations of $X$ lead to changes in the resistance: $\Delta R(t) = (dR/dX)\, \Delta X(t)$. Assuming the current $J$ is constant, then fluctuations of the voltage $V$ is $\Delta V(t) = J\, \Delta R(t)$. In the frequency domain, the spectral density of voltage fluctuations is $S_V(f) = (\Delta V)_f^{\,2} = J^2 (\Delta R)_f^{\,2}$.

To write an equation for $X(t)$, the phenomenological approach developed by L.D. Landau et al. can be taken.[26] This approach is based on the existence of a system with two sets of strongly distinct characteristic times: due to short relaxation times the system reaches so-called "incomplete equilibrium", then it slowly ("quasistationary") relaxes during a longer period of time. The Lorentz fluctuator can be introduced for this situation. According to the above analysis, the system under consideration possesses very distinct characteristic times and we can apply the approach of L.D. Landau et al.[26] and introduce the following equation for the configuration coordinate, $X(t)$,

$$dX/dt - \lambda X = \xi(t). \tag{5}$$

Here, $\lambda$ ($>0$) describes the slow relaxation of $X$, $\xi(t)$ is a ''random force'', the correlator which is proportional to $\lambda$, $<\xi(0)\,\xi(t)> = 2\lambda <X^2>\delta(t)$ with $<X^2>$ being total thermal fluctuation of the quantity $X$. From these equations we readily obtain the spectral density of voltage fluctuations in the Lorentz form:

$$S_V(f) = 2J^2 \left[\frac{dR}{dX}\right]^2 \frac{<X^2>}{\pi f_0} \frac{1}{1+(f/f_0)^2} = \frac{B/f_0}{1+(f/f_0)^2} \tag{6}$$

where $f_0 = \lambda/(2\pi)$. Since non-equilibrium processes (i.e. currents) are responsible for the long times and $\lambda$ describes the rate of relaxation, we can assume that $\lambda$ is proportional to $J$. Then, the proposed model predicts that $f_0 \sim J$ and $S_V(f) \sim J$.

To prove these conclusions we measured the noise for lock-in state at different currents. We found that both the plateau of Lorentzian-shape noise component, $S_{V0}^{LS}$, and the characteristic frequency, $f_0$, are dependent linearly on current, while normalized power noise spectral density of the flicker noise component, $f S_I^{FL}$, has a quadratic dependence on current in molecule-free and molecule-containing junctions (Fig. 7b). Thus, experimental results are in good agreement with theoretical predictions.

It should be noted that $1/f^2$ type of noise was reported in Ref. 16. However, in the paper the frequency dependences of flicker noise, as well as the current-voltage characteristics demonstrated significant sample to sample variation. Therefore only several assumptions without physical interpretation were done there. In contrast, our samples under test showed very reproducible behavior. High stability, up to 50 minutes, in the look-in state was achieved. All fifty samples studied show Lorentzian- shape noise component when the molecule bridges the electrodes. This allows us to state that revealed low-frequency $1/f^2$-noise component in the case of a single molecule junction is the fundamental characteristics of the system. The developed phenomenological model, that takes into account a correlation between charge transfer via a single molecule and structural changes in the coupling, allow us to describe the Lorentzian-noise component of a single molecule locked between two gold electrodes.

## V. CONCLUSION

We measured and compared the electronic noise of improved performance molecule-free and single-molecule-containing mechanically controllable break junctions. Both junctions demonstrated $1/f$ noise characteristics scaling with the square of the current, which is typical for the expected noise increase with the resistance in linear regime. The most striking feature of noise is an additional $1/f^2$ noise component clearly revealed in all MSM junctions only in the case when a single molecule bridged the nanoelectrodes. In the examined set of molecule containing samples a well defined time constant at certain small current was extracted. It was found that the time constant has linear dependence as a function of current in the small current regime. The recorded $1/f^2$ electric noise component relative to a single bridging molecule is interpreted as a manifestation of a dynamic reconfiguration of molecular coupling to the metal electrodes during current flow. The reconfiguration changes occur without complete bond breaking and involve near-configuration states with very similar electric properties. Using the fact of the existence of strongly distinct characteristic times in the system, we developed a phenomenological model that relates the charge transfer via a single molecule with its reconfiguration and characteristic time of a single molecule. Thus we interpret the revealed Lorentzian-type noise as the manifestation of a system reconfiguration under the charge transfer via a single molecule. Obtained experimental results, particularly dependencies of the noise amplitude and the characteristic noise frequency on the current, support this interpretation. The current and noise models are in good agreement with experimental data obtained in the investigated range of small currents.

The results obtained provide a noise-based route to study characteristic times in molecule containing break junctions and highlight opportunities for detailed analysis of microscopic mechanism of charge transfer in systems with individual electrically addressed molecules. The results should be taken into account for the development of devices for molecular electronics.

## ACKNOWLEDGMENTS

Financial support from Helmholtz-CSC is gratefully acknowledged.